\documentclass[aps,reprint,prb,superscriptaddress]{revtex4-1}
\pdfoutput=1
\usepackage{filecontents} 
\usepackage{hyperref} 
\usepackage{amsmath} 
\usepackage{verbatim} 
\usepackage{braket} 
\usepackage{array} 
\usepackage[percent]{overpic} 
\usepackage[title]{appendix} 

\begin{document}

\title{Deviation from the dipole-ice model in the new spinel spin-ice candidate, MgEr$_2$Se$_4$}

\author{D. Reig-i-Plessis}
\affiliation{Department of Physics and Seitz Materials Research Laboratory, University of Illinois at Urbana-Champaign, Urbana, Illinois, 61801, USA}
\author{S. van Geldern}
\affiliation{Department of Physics and Seitz Materials Research Laboratory, University of Illinois at Urbana-Champaign, Urbana, Illinois, 61801, USA}
\author{A. A. Aczel}
\affiliation{Neutron Scattering Division, Oak Ridge National Laboratory, Oak Ridge, Tennessee, 37831, USA}
\author{D. Kochkov}
\affiliation{Department of Physics and Seitz Materials Research Laboratory, University of Illinois at Urbana-Champaign, Urbana, Illinois, 61801, USA}
\author{B. K. Clark}
\affiliation{Department of Physics and Seitz Materials Research Laboratory, University of Illinois at Urbana-Champaign, Urbana, Illinois, 61801, USA}
\author{G. J. MacDougall}
\email{gmacdoug@illinois.edu}
\affiliation{Department of Physics and Seitz Materials Research Laboratory, University of Illinois at Urbana-Champaign, Urbana, Illinois, 61801, USA}

\begin{abstract}
In spin ice research, small variations in structure or interactions drive a multitude of different behaviors, yet the collection of known materials relies heavily on the `227' pyrochlore structure.
Here, we present thermodynamic, structural and inelastic neutron scattering data on a new spin-ice material, MgEr$_2$Se$_4$, which contributes to the relatively under-explored family of rare-earth spinel chalcogenides.
X-ray and neutron diffraction confirm a normal spinel structure, and places Er$^{3+}$ moments on an ideal pyrochlore sublattice.
Measurement of crystal electric field excitations with inelastic neutron scattering confirms that the moments have perfect Ising character, and further identifies the ground state Kramers doublet as having dipolar-octupolar form with a significant multipolar character.
Heat capacity and magnetic neutron diffuse scattering  have ice-like features, but are inconsistent with Monte Carlo simulations of the nearest-neighbor and next-nearest-neighbor dipolar spin-ice (DSI) models.
A significant remnant entropy is observed as $T\rightarrow0$~K, but again falls short of the full Pauling expectation for DSI, unless significant disorder is added.
We show that these observations are fully in-line with what is recently reported for CdEr$_2$Se$_4$, and point to the importance of quantum fluctuations in these materials.

\end{abstract}

\maketitle

\section{Introduction}

One of the most notable aspects of the spin-ice class of compounds is the wide range of interesting behaviors they exhibit.
Originally of interest as a magnetic analogue of water ice and a playground for thermodynamic models\cite{harris1997, harris1998, bramwell1998, moessner1998, bramwell2001, ramirez1999}, the field has expanded dramatically to include an assortment of interesting variations, including Kagom\`e ice\cite{tabata2006,fennell2007}, ordered spin ice\cite{mirebeau2005}, dynamic spin ice\cite{zhou2008} and quantum spin ice (QSI)\cite{gingras2014,savary2017,sibille2018experimental}.
The QSI materials are prime candidates for a U(1) quantum spin liquid (QSL)\cite{huse2003,hermele2004, savary2012, lee2012, gingras2014, hao2014, petrova2015, shannon2012,chen2016, benton2012}, which itself may have multiple variations\cite{taillefumier2017} including distinct, symmetry-enriched phases\cite{huang2014,li2017}.

The unifying feature in this panoply is the underlying ``classical'' spin-ice model, wherein spins on a pyrochlore lattice with local $\langle 111 \rangle$ Ising anisotropy and ferromagnetic interactions freeze into an extensively degenerate ``ice'' phase, characterized by a 2-in-2-out (TITO) constraint on the constituent tetrahedra.
This TITO constraint famously maps onto the Bernal-Fowler ice rules for proton-oxygen bond lengths in frozen water\cite{bernal1933}, and the associated remnant `Pauling entropy' as $T^*~\rightarrow~0~\text{K}$\cite{pauling1935,ramirez1999} remains the primary experimental signature of a classical spin-ice state.
The TITO constraint can further be mapped to a divergence-free flux, allowing one to reinterpret the ice as a ``magnetic Coulomb'' phase\cite{henley2010} wherein thermodynamic properties can be calculated by considering a gas of magnetic monopoles\cite{castelnovo2008, morris2009,kadowaki2009}.

In real spin ice materials, the local Ising condition is a result of trigonal crystal electric fields (CEF), and effective ferromagnetic interactions emerge from summing nearest-neighbor exchange and dipole terms\cite{bramwell2001,gardner2010}.
This dipolar spin-ice (DSI) model\cite{denhertog2000,melko2004,isakov2005} is sufficient to explain the origin of the ice state, and has been successful in reproducing measured heat capacity\cite{zhou2012} (HC) and basic features of neutron diffuse scattering patterns\cite{fennell2007,morris2009,chang2010,arnab2013} in known classical spin ices; this includes Ho$_2$Ti$_2$O$_7$\cite{harris1997,Fennell09Ho2Ti2O7}, Dy$_2$Ti$_2$O$_7$\cite{ramirez1999}, and associated stannates ($R_2$Sn$_2$O$_7$)\cite{ke2007,ke2008} and germanates ($R_2$Ge$_2$O$_7$)\cite{zhou2011, hallas2012}.
The breadth of behaviors described above, however, is a testament to the importance of further degeneracy breaking terms.
Further neighbor exchange is needed to explain details of diffuse scattering and reproduce measured critical fields\cite{ruff2005, melko2004}.
Quantum fluctuations result from either transverse molecular fields\cite{tomasello2015} or from multipolar superexchange interactions\cite{onoda2011,rau2015,iwahara2015}, with the latter invoked to explain experimental data in Pr$_2$Sn$_2$O$_7$ and Pr$_2$Zr$_2$O$_7$\cite{onoda2010, kimura2013, petit2016_2,sibille2016}.
The unique dipolar-octupolar (DO) character of moments in materials such as Nd$_2$Zr$_2$O$_7$\cite{xu2016,huang2014,15_lhotel_fluctuations_DO_Nd2Zr2O7} and Ce$_2$Sn$_2$O$_7$\cite{sibille2015} is linked to the possibility of symmetry-enriched QSL phases\cite{huang2014,li2017}.

There is thus clear motivation to extend the study of spin-ice physics to materials beyond the 227 oxides, with different variations in local structure and interactions.
The cubic spinels (AB$_2$X$_4$) are prime candidates, as they share the same Fd$\bar{3}$m space group and pyrochlore sublattice as the 227 compounds, but differ in the octahedral arrangement of local chalcogen anions about B-site spin positions\cite{lee2010, takagi_book, radaelli2004}.
Sizable trigonal CEFs create $\langle 111 \rangle$ easy axes on some B-site ions, which play a defining role for material properties\cite{mun2014,macdougall2012,ehlers2012}.
Ferromagnetically coupled $\langle 111 \rangle$ easy axis spins reminiscent of spin ices have been reported in several spinels leading to frustration observed through diffuse scattering in single crystals \cite{tomiyasu2011}, or leading to two-in-two-out ordered states of the B-site sublattice \cite{macdougall2012,brodsky2014, ma2015,lee2017}.
In the singular system, CdEr$_2$Se$_4$, remnant Pauling entropy has been reported\cite{lago2010}, and a very recent study on the same material has claimed DSI-like spin-correlations and an anomalously fast monopole hopping rate\cite{gao2018}.

Here, we present data on a new spinel, MgEr$_2$Se$_4$, which provides another interesting counterpart to known spin-ice materials.
We provide x-ray (XRD) and neutron powder diffraction (NPD) data which confirm an ideal pyrochlore sublattice of Er$^{3+}$ moments, but with a cubic lattice parameter $\approx10\%$ larger than Dy$_2$Ti$_2$O$_7$.
Inelastic neutron scattering (INS) data reveal that the moments have ideal Ising anisotropy, and further show that they have a significant multipolar character, in fact demonstrating the characteristic DO symmetry\cite{huang2014}.
Both heat capacity and magnetic diffuse scattering data exhibit qualitative features of classical spin ice correlations.
Follow-up Monte Carlo (MC) simulations, however, show that the collective data are inconsistent with nearest-neighbor and next-nearest neighbor DSI models.
This may be a natural consequence of the multipolar character of the Er$^{3+}$ moments, which seed significant quantum fluctuations.

\section{Sample preparation and crystal structure}

Polycrystalline samples of MgEr$_2$Se$_4$ were prepared via a two-step solid state reaction, following the method described by Flahaut \cite{flahaut1965}.
The precursors MgSe and Er$_2$Se$_3$ were prepared by the direct reaction of stoichiometric amounts of the elements at $650^{\circ}$C.
Stoichiometric quantities of the two precursors were then combined, pelletized and reacted in vacuum at $1000^{\circ}$C for two days, this step was repeated at least one more time for the precursors to fully react.
Structure and purity were confirmed using a PANalytical X'Pert$^3$ X-ray powder diffractometer at the Center for Nanophase Materials Sciences at Oak Ridge National Laboratory (ORNL).
NPD measurements were performed with the HB-2A powder diffractometer at ORNL's High Flux Isotope Reactor, using a 3.6~g sample and neutron wavelengths of $\lambda = 1.54~\text{\AA}$ and $\lambda = 2.41~\text{\AA}$, with collimators open-21$^\prime$-12$^\prime$ and open-open-12$^\prime$, respectively.
Structural refinements were performed using the FULLPROF \cite{rodriguez1993} software suite.
Additional measurements in a magnetic field used the CTAX instrument with $\lambda=5~\text{\AA}$ neutrons.
INS was performed with the SEQUOIA \cite{Granroth20061104} fine-resolution Fermi chopper spectrometer at ORNL's Spallation Neutron Source (SNS).
Measurements were collected with incident energies $E_i = 30~\text{meV}$ and $E_i = 50~\text{meV}$ with the fine Fermi chopper spinning at frequencies of 300~Hz and 360~Hz respectively.
Magnetization and specific heat measurements were performed in the Seitz Materials Research Laboratory at Illinois using a Quantum Design MPMS3 and PPMS, respectively.

\begin{table}[htb] 
 \begin{ruledtabular}
   \begin{tabular}{ l c c c }
    \multicolumn{4}{c}{MgEr$_2$Se$_4$ lattice parameters} \\
    \multicolumn{4}{c}{space group Fd$\bar{3}$m}\\

     & XRD 300K & NPD 38K & NPD 470mK \\
    \hline
    a & 11.5207(14) & 11.4999(42) & 11.5048(81) \\
    $\chi^{2}$ & 10.41 & 6.39 & 8.63 \\
    $\chi^{2}$ Lebail & 11.40 & 6.83 & 9.53 \\
    Se deficiency (\%) & 0.00(70) & 0.00(98) & 0.0(1.2) \\
    Site inversion (\%) & 0.00(47) & 0.0(3.7) & 0.0(4.5) \\

  \end{tabular}

  \begin{tabular}{  l  c  c  c  c }
    \multicolumn{5}{c}{MgEr$_2$Se$_4$ atom positions} \\ 
      & x & y & z & $B_{\text{iso}} ($\AA$^2)$ \\ 
    \hline
    Mg & 0.375 & 0.375 & 0.375 & 0.2(1) \\ 
    Er & 0.000 & 0.000 & 0.000  & 0.39(5)\\
    Se & 0.2456(9) & 0.2456(9) & 0.2456(9) & 0.40(3) \\

  \end{tabular}

 \caption{Structural parameters obtained from the XRD and NPD refinements of MgEr$_2$Se$_4$ data.
 The structural parameters in the lower part of the table are from the 38K NPD refinement.}
 \label{tab:struct}

\end{ruledtabular}
\end{table}

\begin{figure}[hbt]
 \centering
 \includegraphics[width=\columnwidth]{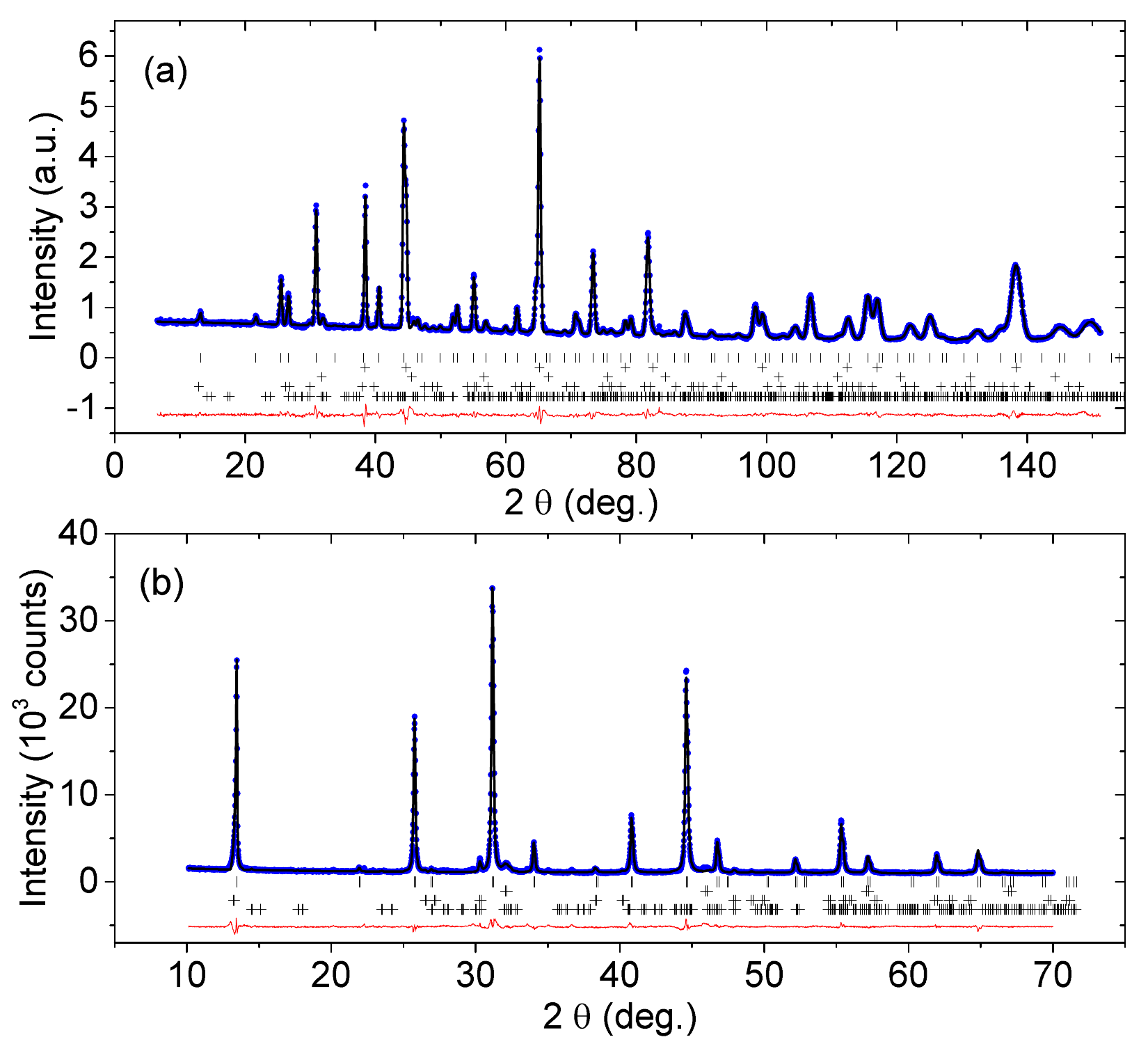}
 \caption{NPD (top) and XRD (bottom) data for the primary MgEr$_2$Se$_4$ sample with data points in blue, best fit Rietveld refinement in black, and the difference shown in red. Tick marks show positions of MgEr$_2$Se$_4$ peaks, while crosses show position of peaks from fit impurity phases.}
 \label{fig:diffraction}
\end{figure}

A large volume sample of MgEr$_2$Se$_4$ was prepared for exploration with neutron scattering and, unless stated otherwise, was used to obtain all data presented in the figures below.
Purity and structure were studied with both x-ray diffraction (XRD) and neutron powder diffraction (NPD), and diffraction patterns on our primary sample are shown in Fig.~\ref{fig:diffraction}, along with the results of FULLPROF refinements.
Impurity peaks in both patterns are denoted by crosses, and were largely accounted for by the orthorhombic phase of Er$_2$Se$_3$ (1.8--5.0~\%), which is consistent with a small amount of Mg evaporating during synthesis.
In addition to Er$_2$Se$_3$, refinements also showed small amounts of Er$_2$O$_2$Se (2.3--2.5~\%) and elemental Er (1.2~\%), as well as some small unindexed impurity peaks which are not consistent with any known compound.
The best fit refinement implies that the sample had purity of  94.66(62)~\% and 91.4(1.5)~\% by mass from the XRD and NPD fits respectively.
A model independent estimate of the impurity fraction of 7.8(1.2)~\% by mass was obtained by comparing the integrated intensity of XRD Bragg peaks associated with the majority phase and the sum intensity of everything else above background. The weighted average of the three estimates gives a value of 6.2(1.2)~\% impurity phase in the sample.

\begin{figure*}[hbt] 
 \centering
 \includegraphics[width=\linewidth]{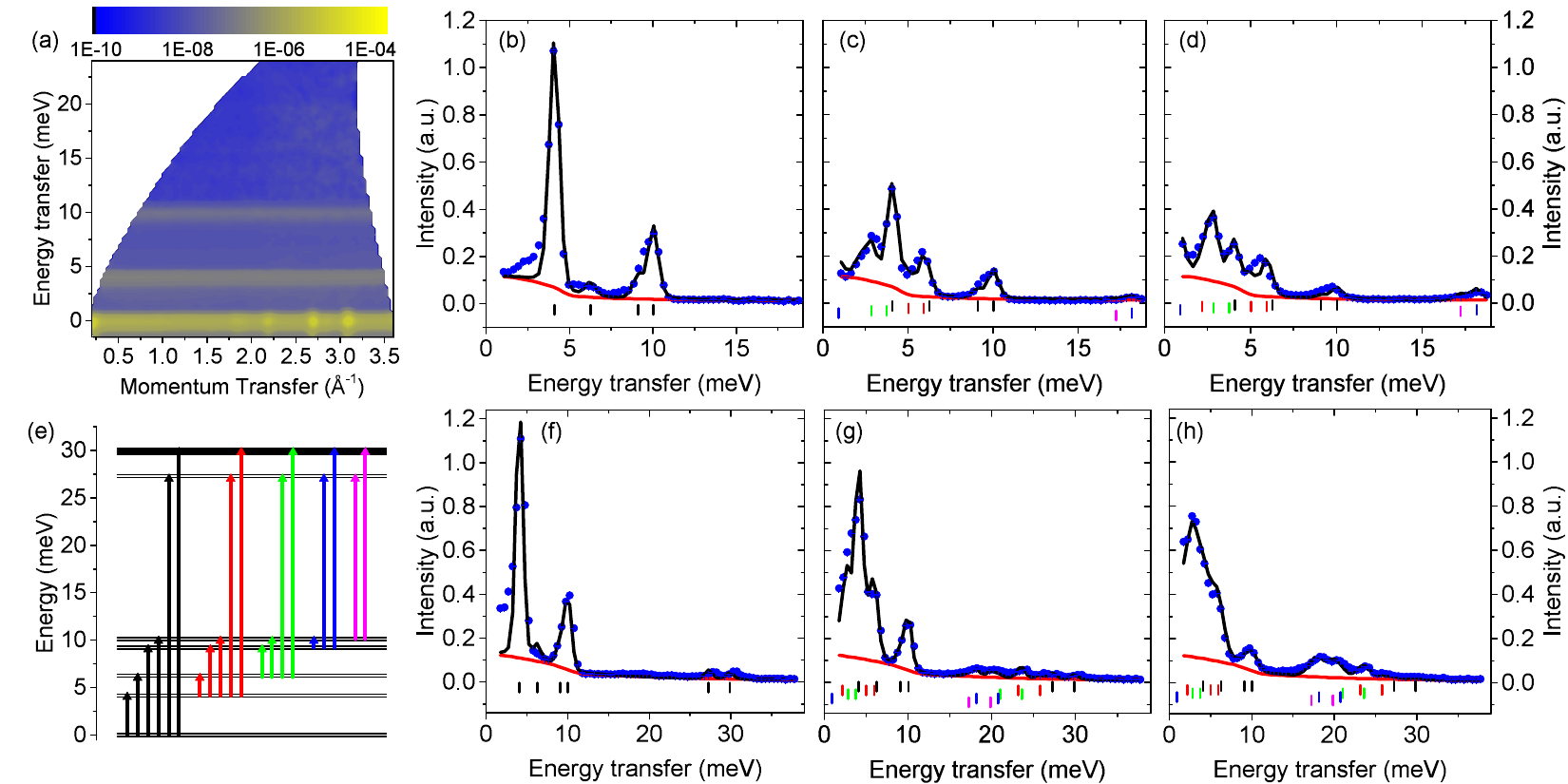}
 \caption{ Representative INS data, including a false color plot of scattering intensity in the energy-momentum plane at temperature $T = 5$~K and $E_i = 30$~meV (a), and associated cuts at momentum $|Q| = [2, 2.5]$~\AA$^{-1}$ at incident energy $E_i = 30$~meV and $E_i = 50$~meV for (b)-(d) and (f)-(h) respectively. Temperatures are $T$ = 5, 40 and 150 K from left to right in both (b)-(d) and (f)-(h). Solid lines represent the best fits described in the text, blue dots represent the INS data with error bars too small to be visible. Colored marks denote positions of transitions, color coded by initial occupied level.
 (e) The crystal field energy scheme inferred from the above fitting, with color coded arrows denoting observed transitions.
}
 \label{fig:seq}
\end{figure*}

Both NPD and XRD show that the MgEr$_2$Se$_4$ phase is of high quality and shows no observable defects in the structure.
To test for any structural defects, we allowed Se occupancy, Mg occupancy as well as Er and Mg site inversion to vary.
Results of best fits are shown in Table~\ref{tab:struct} and show no such defects with bounds of $<1~\%$.

As an additional model independent check for point defects, we performed Le Bail refinements \cite{lebail1988} of our data and compared the $\chi^2$ of those fits to our best Rietveld refinement.
In Lebail fits, the structure factor is not calculated, and instead every peak height is allowed to vary and are fit independently -- effectively identifying the ideal description for peaks associated with a single phase in a mixed powder\cite{toby2006}.
In the current case, we see that the $\chi^2$ achieved through a Le Bail peak-by-peak fitting of the majority phase is no smaller than that achieved via the above Rietveld refinement.
This is a powerful result, which effectively eliminates the existence of cation inversion, off-stoichiometry on the Se-sublattice, or any other point defect which has the capacity to change the height in a neutron scattering pattern.

\begin{figure}[hbt] 
 \centering
\includegraphics[width=\columnwidth]{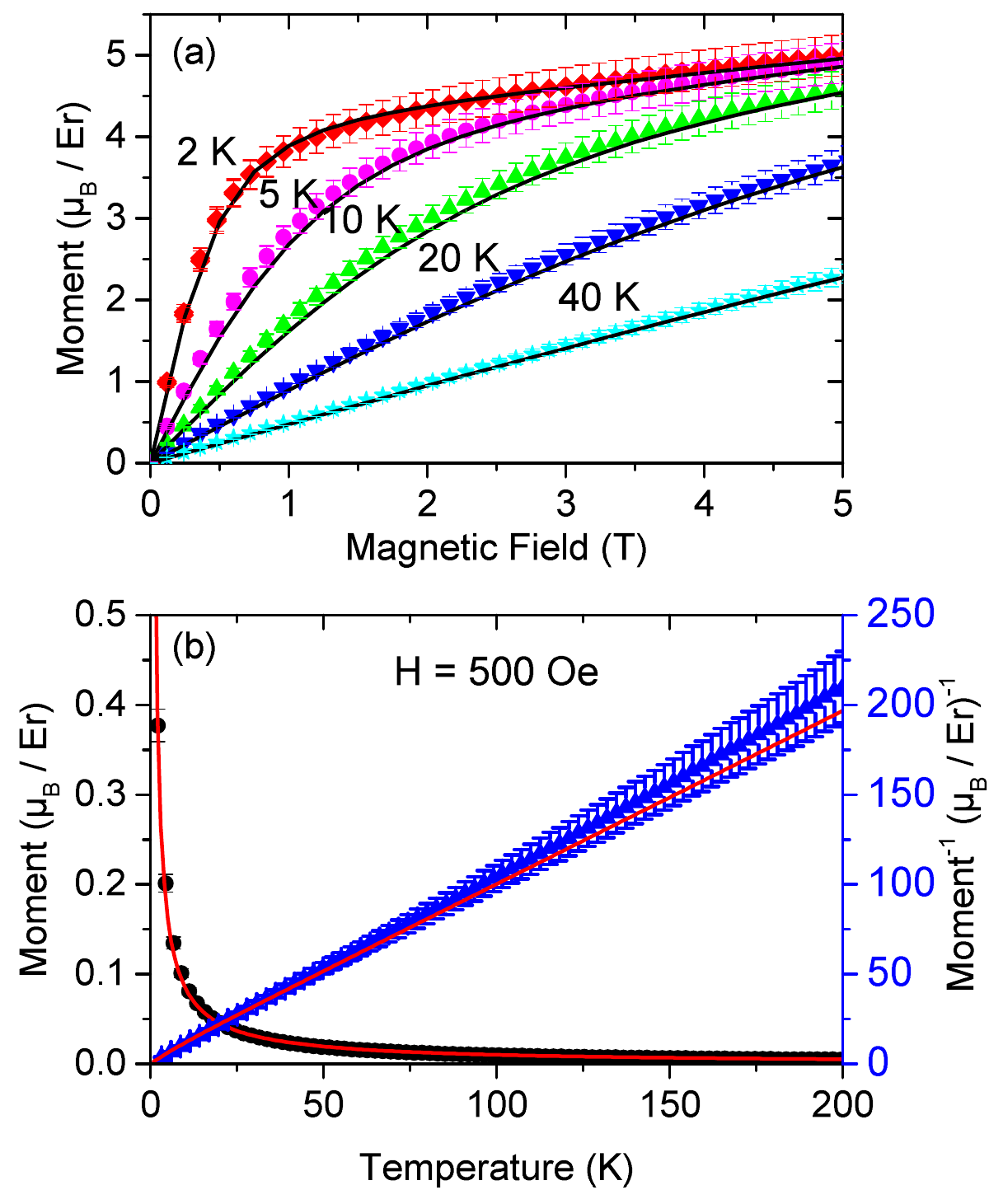}
\caption{ Measured magnetization of a powder sample of MgEr$_2$Se$_4$ as a function of field (a) for T = 2, 5, 10, 20 and 40 K and as a function of temperature (b). Solid lines are the calculated magnetization at the same temperatures based on the CEF parameters found in this paper.
}
 \label{fig:mag}
\end{figure}

\section{Inelastic neutron scattering}

The local CEF environment of Er$^{3+}$ was determined with INS. In Fig.~\ref{fig:seq}(a), we show a representative INS data set collected using $E_i = 30$~meV at $T = 5$~K, wherein CEF excitations out of the ground state appear as bright dispersionless modes near $E \approx 4$ meV and $E \approx 10$~meV.
To access transitions form higher energy levels additional measurements were performed at $T = 40$~K and $T = 150$~K, the same temperatures were also measured at $E_i = 50$~meV.
In Figs.~\ref{fig:seq}(b--d) and (f--h), we show cuts obtained by integrating the INS data over momentum interval $Q = [2,2.5]\text{\AA}^{-1}$, chosen to maximize the available energy range.
A background contribution (shown as a red line) is interpolated from hand picked points at energy transfers away from CEF peak positions.
The scattering intensities of the six data sets were fit simultaneously to expectations from the model crystal field Hamiltonian
\begin{equation}
H = \sum_{nm} B_n^m O_n^m,
\label{eq:CEF}
\end{equation}
where $O_n^m$ are the Stevens' operators\cite{stevens1952}.
Fits were performed using a mix of random walk grid search and gradient search methods explained in detail in Appendix \ref{appendix:INS_fit} and began with an initial guess calculated using a point charge model\cite{hutchings1964} and the known structure.
The CEF levels and electron wavefunctions were calculated with the quantization axis along the local $\langle 1 1 1 \rangle$ directions, with only coefficients $B_2^0, B_4^0, B_4^3, B_6^0, B_6^3, B_6^6$ which can be non-zero by symmetry.
Predicted peaks were convoluted with a Voigt profile to account for instrument resolution.
Through simultaneous consideration of nearly two dozen observed peaks, we determined the six most likely CEF parameters for MgEr$_2$Se$_4$ to be (in meV): $
B_2^0 = -4.214(63)\times10^{-2},
B_4^0 = -6.036(30)\times10^{-4},
B_4^3 = -1.3565(67)\times10^{-2},
B_6^0 = 3.264(16)\times10^{-6},
B_6^3 = -3.791(75)\times10^{-5} \text{and},
B_6^6 = 2.194(65)\times10^{-5}$. The scattering pattern associated with these parameters is denoted by solid lines in Figs.~\ref{fig:seq}(b--d) and (f--h), and successfully reproduces both intensity and position of all considered modes and predicts no errant peaks.
The data further reveals the existence of two small peaks at energies $E\approx 2$~meV and $E\approx 5$~meV, with spectral weight consistent with the $\approx7\%$ impurity in the sample.

Figure~\ref{fig:seq}(e) shows the crystal field levels calculated from the above $B_n^m$, along with the various transitions observed in our scattering data.
The associated wavefunctions are given in their entirety in Appendix~\ref{appendix:INS_fit}, and the ground state Kramers doublet is
\begin{multline*}
 \ket{\psi_0^\pm} = \pm0.9165(7)\ket{\pm15/2} + 0.3600(11)\ket{\pm9/2} \\
 \pm 0.1581(16)\ket{\pm3/2} -  0.0731(15)\ket{\mp3/2} \\
 \pm 0.0036(7)\ket{\mp9/2} + 0.0035(14)\ket{\mp15/2},
\end{multline*}

which implies perfectly Ising spins
\begin{equation*}
\bra{\psi_0^+}J_x\ket{\psi_0^-} = \bra{\psi_0^+}J_y\ket{\psi_0^-} =0
\end{equation*}
with moment $m_z = g_J\bra{\psi_0^+}J_z\ket{\psi_0^+} =  8.3(1) \mu_B$, where the Land\'e g-factor $g_J=\frac{6}{5}$ for Er$^{3+}$.
This wavefunction also facilitates significant non-dipolar superexchange interactions, as discussed below.
The two lowest excited doublets are at energies $E_1 = 4.02(2)~\text{meV}$ and $E_2 = 6.40(2)~\text{meV}$, significantly larger than interaction energies determined below, but still low enough to impact thermodynamic properties at temperatures $T > 5$~K.

\section{Magnetization}

To check the validity of the CEF fits, inferred levels were used to calculate the magnetization in the paramagnetic phase for a range of applied fields.
For the case of low lying CEF excitations, the effect of mixing of excited CEF levels must be taken into account for calculating the magnetization \cite{Bonville_2013}.

Magnetization was obtained using a non-interacting model with total Hamiltonian of the $J = 15/2$ Er$^{+3}$ multiplet CEF plus the Zeeman energies
\begin{equation}
H = H_\text{CEF} + H_Z.
\label{eq:in_field_potential}
\end{equation}
The CEF Hamiltonian is defined as before in Eq.~\ref{eq:CEF}
and the Zeeman term
$H_Z = - g_J \mathbf{H} \cdot \hat J $.
For powder averaging, the magnetization was calculated for over 1500 different local applied field directions for each temperature and field value.
Finally Boltzmann statistics were used to find the occupation of each perturbed CEF level and then calculate the associated magnetization.
The results are shown as solid lines in Fig.~\ref{fig:mag}.
As one can see, this calculation largely reproduces the magnitude, temperature and field dependence of measured magnetization with \textit{zero} fit parameters.
Particularly notable, the calculation was able to reproduce the linear field dependence for $H > 2$~T, which we confirmed is the result of the field perturbation of local eigenstates in excited doublets and is not captured in the simplified pseudospin 1/2 models for magnetization that has been used to argue for Ising behavior in past rare earth compounds\cite{bramwell2000, xu2015, sibille2015, hallas2015, anand2016}.
The overall agreement here provides strong support for the inferred levels above and the Ising character of the Er$^{3+}$ moments at temperatures $T<5$~K.
Although we used a non-interacting model, we believe the agreement at $T = 2$~K is to be expected given the small exchange and dipole energies compared to 2 K and the degree of frustration in this material.
Additional detail on the calculation method and a comparison to a pseudospin 1/2 models are given in Appendix~\ref{appendix:mag_calc}.

\begin{figure}[htb] 
 \centering
 \includegraphics[width=\columnwidth]{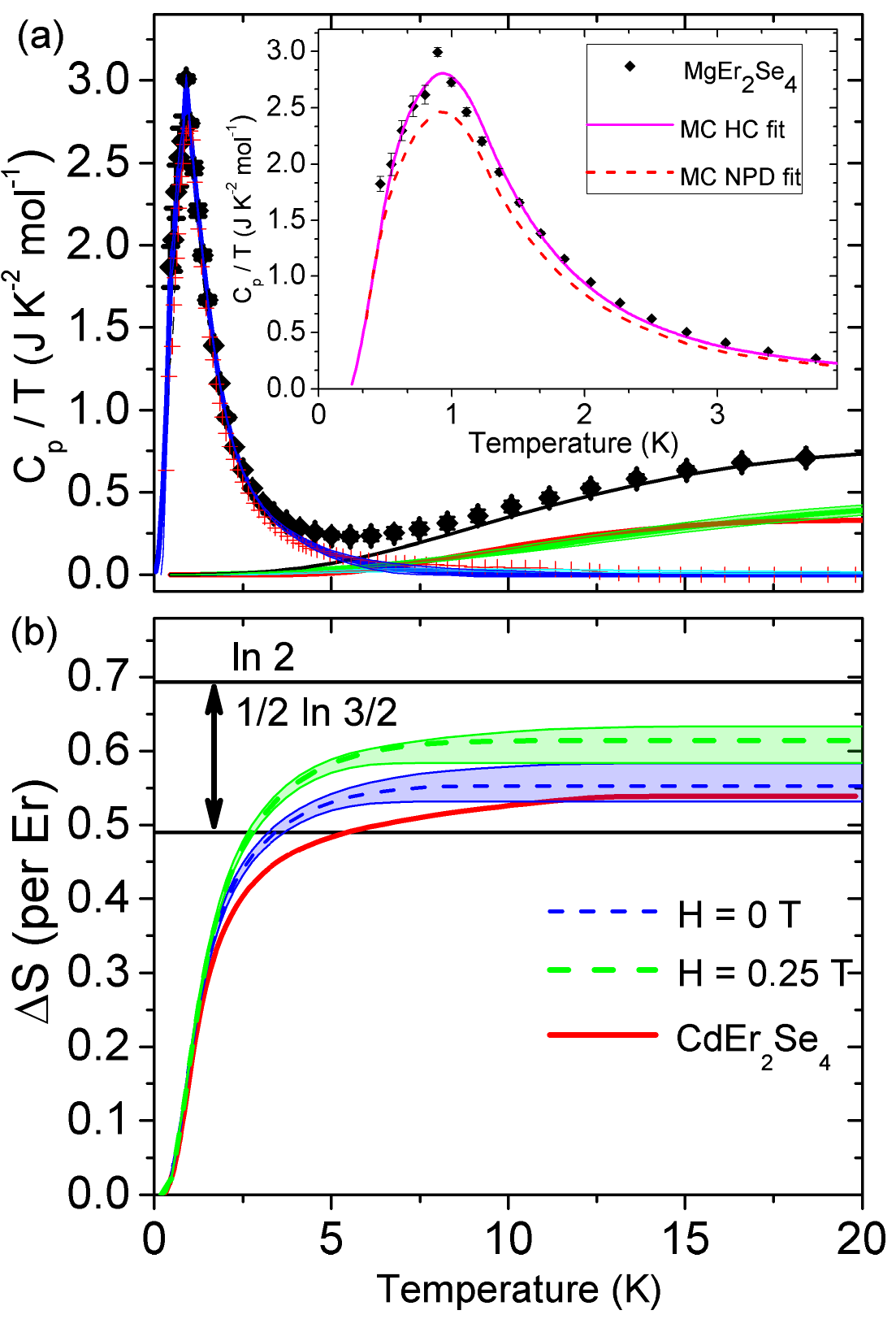}
 \caption{(a) Measured heat capacity of MgEr$_2$Se$_4$ as black diamonds, with lines denoting estimated contributions from phonons (green), crystal fields (red), impurities (cyan) and magnetic degrees-of-freedom (blue). The black line is the sum of phonon, impurity and crystal field contributions. Red crosses show the data from CdEr$_2$Se$_4$. (Inset) Magnetic contribution to heat capacity on a magnified scale, with best fit curves from MC simulations to the HC data (solid line) and to the NPD data (dashed line).  (b) Entropy per Er$^{3+}$ moment from magnetic contribution to heat capacity, for zero (blue) and applied field (green) dashed lines. Shaded regions quantify uncertainty. Data on the material CdEr$_2$Se$_4$ shown as red crosses were taken from Ref.~\onlinecite{lago2010}, and corrected for recently measured\cite{gao2018} crystal field levels.}

\label{fig:HC}
\end{figure}

\section{Heat Capacity}

Measured heat capacity in the range $0.45~\text{K} < T < 20~\text{K}$ is shown in Fig.~\ref{fig:HC}(a), with solid lines representing contributions from phonons, CEFs, impurities and magnetic degrees of freedom, and the linewidth representing the uncertainty.
The CEF contribution was calculated exactly from the multi level partition function given by the CEF scheme determined above, and contains non-trivial contributions from both the E$_1$ and E$_2$ doublets.
The contribution to the heat capacity from the impurity phases was taken into account by subtracting off the heat capacity expected from a system with transition energies at $E\approx 2$~meV and 5~meV, and appropriately scaled to be between 5 and 9~\%; molar mass of the sample was also accordingly scaled.
The energies chosen for the impurity phase come from extra modes observed in INS.
A small amount of an impurity was found to order from the neutron powder diffraction and was included in the error bars of the impurity contribution -- more details can be found in Appendix~\ref{appendix:mag_impurity}.
Phonons were modeled by fitting data in the range $7~\text{K} < T < 25~\text{K}$ to the Debye model after CEF and impurity contributions were subtracted.
The black line in Fig.~\ref{fig:HC}(a) is the sum of these contributions, and is seen to perfectly describe data above 10 K.
The remaining contribution was entirely attributed to the Er$^{3+}$ moments on the pyrochlore sublattice.
This Er$^{+3}$ magnetic contribution is dominated by a single broad peak near $T^* \approx 1.1~\text{K}$, which has a height and position broadly consistent with expectations for the DSI model\cite{denhertog2000}.
A similar analysis was done with the heat capacity data of CdEr$_2$Se$_4$ taken from Ref.~\onlinecite{lago2010}, after suitably updating the treatment of CEF energy levels using recently measured values taken from Ref.~\onlinecite{gao2018}.
We plot the resulting magnetic contribution in CdEr$_2$Se$_4$ as red crosses in Fig.~\ref{fig:HC}(a) and the corresponding calculated entropy as a solid red line in Fig.~\ref{fig:HC}(b).
We find both the heat capacity and residual entropy of the two erbium selenium spinels to be remarkably similar, adding to confidence in the data and analysis.

To gain further insight into the exact temperature dependence of the heat capacity, we performed MC simulations on a 2048-site cluster with periodic boundary conditions in all directions, as described in Appendix~\ref{appendix:MC}.
This simulation used the Hamiltonian:
\begin{multline*}
H = -3J_{nn}\sum_{\langle i,j \rangle}\mathbf{S}_i^{z_i}\cdot\mathbf{S}_j^{z_j}
-3J_{nnn}\sum_{\langle\langle i,j \rangle\rangle}\mathbf{S}_i^{z_i}\cdot\mathbf{S}_j^{z_j}\\
+ \frac{3D_{nn}}{5}r_{nn}^3\sum_{i<j}\left(\frac{\mathbf{S}_i^{z_i}\cdot\mathbf{S}_j^{z_j}}{|\mathbf{r}_{ij}|^3} - \frac{3(\mathbf{S}_i^{z_i}\cdot\mathbf{r}_{ij})(\mathbf{S}_j^{z_j}\cdot\mathbf{r}_{ij})}{|\mathbf{r}_{ij}|^5}\right),
\end{multline*}
where the parameter $J_{nn}$ ($J_{nnn}$) represents nearest (next-nearest) neighbor exchange interactions, and the strength of the dipole interaction was fixed to $D_{nn} = 1.06$~K, as determined from the measured structure.
For $J_{nnn} = 0$, we confirmed that our calculations match previously published results\cite{denhertog2000}.
We found that the primary effect of low $J_{nnn}$ was to symmetrically broaden and increase the height of the peak in heat capacity, regardless of sign.
As discussed in the Appendix~\ref{appendix:MC}, sufficiently large $J_{nnn}$ are seen to drive the system to either a $Q=0$ or $Q=X$ ordered state for ferromagnetic or antiferromagnetic interactions, respectively.

The best fit of our magnetic heat capacity data gave values $J_{nn} = 0.06$~K and $J_{nnn} = -0.1$~K, producing the curve shown in the inset of Fig.~\ref{fig:HC}(a). This fit described the data adequately (reduced $\chi^2 = 16.7$), and significantly better ($\chi^2 = 68.8$) than the curve expected using exchange parameters estimated from fits of NPD data, $J_{nn} = 0.06$~K and $J_{nnn} = 0$~K, discussed in more detail below. In fact, though small on the scale of $D_{nn}$, we note that our MC calculations were found to be entirely incompatible with the HC data without assuming a ferromagnetic $J_{nn}$ and setting $\left| J_{nnn}/J_{nn}\right| > 1$. Both observations stand in contrast to known 227 classical ice systems, but the former may be consistent with the near $90^\circ$ Er-Se-Er superexchange path between nearest neighbors in the spinel structure. 
As discussed below, the sizable value for J$_{nnn}$ is inconsistent with our magnetic diffuse scattering data and may point to other relevant physics.

In Fig.~\ref{fig:HC}(b), we plot the change in entropy from 0~K obtained from the integrated magnetic heat capacity from data in fields $H = 0$~T and $H = 0.25$~T, where MC data was used to extrapolate below the first data point at $T = 0.45$~K . The relatively small field of $H=0.25$~T was chosen in order to minimize changes to the low lying CEF levels and the broadening of those peaks in the HC data, allowing consistent analysis of data taken both in and out of field.
Shading represents experimental uncertainty, which is dominated by uncertainty in the impurity volume fraction. The $H = 0$~T data reveal a sizable residual entropy which is partially relieved with small applied fields, broadly consistent with spin-ice behavior\cite{ramirez1999}, but significantly less than the value of $1/2 \ln \left( 3/2 \right)$ predicted for the DSI model\cite{pauling1935,ramirez1999}.
This is true for both the current data, but also the data taken from Ref~\onlinecite{lago2010} on related material CdEr$_2$Se$_4$ after correcting for contributions from subsequently measured CEF excitations measured recently\cite{gao2018}. Intriguingly, one sees that the remnant entropy of both systems approach the same value, but fall far short of the full Pauling value. This agreement despite the differing level of purity in the two materials\cite{lago2010,gao2018} builds confidence that the data reflect intrinsic physics. The deviation from Pauling entropy implies that some TITO spin configurations are being removed from the degenerate manifold by an interaction term outside the DSI model.

\begin{figure*}[htb] 
 \centering
 \includegraphics[width=1.8\columnwidth]{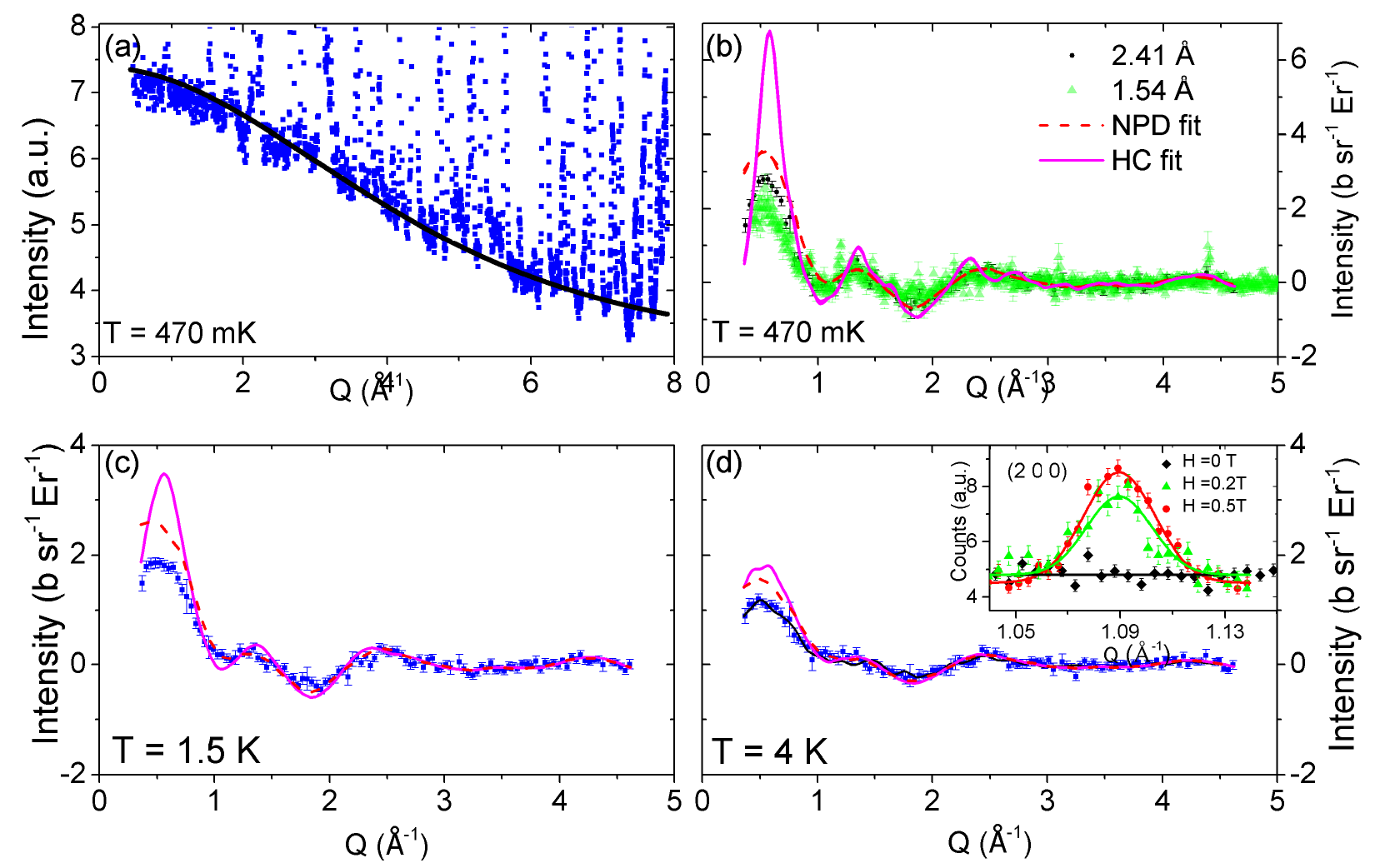}
 \caption{ (a) Diffuse NPD data at $T = 38$~K, along with a fit (solid line) to the ideal form factor for Er$^{3+}$ spins. Panels (b), (c) and (d) respectively show the difference between the low temperature scattering ($T = 470$~mK, 1.5~K and 4~K) and data at $T = 38$~K. Solid lines are intensity from representative snapshots of Monte Carlo configurations using the parameters for $J_{nn}$ and $J_{nnn}$ which best fit the heat capacity data. Dashed lines in (b), (c) and (d) are from a snapshot of Monte Carlo best fit to the NPD data. The inset of (d) shows scattering at the (2 0 0) position in zero and small applied fields. }
 \label{fig:hb2a}
\end{figure*}

Though consideration of the solid curve in Fig.~\ref{fig:HC}(a) might imply the degeneracy breaking term could be $J_{nnn}$, this conclusion is not supported by our NPD measurements of magnetic diffuse correlations, as discussed below.

\section{Neutron powder diffraction}

A summary of our NPD results is presented in Fig.~\ref{fig:hb2a}.
In Fig.~\ref{fig:hb2a}(a), we show that the diffuse background at $T = 38$~K is dominated by paramagnetic scattering, which fit well to Er$^{3+}$ form factor squared and was used to normalize subsequent data.
The paramagnetic scattering was then used to plot data at lower temperatures on an absolute intensity scale- an important step in the analysis of the diffuse scattering, since MC simulations reveal that much of the relevant information regarding next-nearest neighbor interactions is encoded in the scattering intensity at particular values of $Q$.
In panels (b)-(d), we plot on an absolute scale the scattering data taken with neutron wavelength $\lambda=2.41$~\AA~ at $T = 0.47$~K, 1.5~K and 4~K, with the contribution at $T = 38$~K subtracted to isolate the magnetic contribution.
Low temperature magnetic correlations are largely short-ranged and consistent with an ice-like state.
A handful of Bragg peaks were observed in the $T = 0.47$~K pattern only, reliably associated with the small impurity fraction and subtracted from the pattern in Fig.~\ref{fig:hb2a}(b).
A second data set taken with $\lambda=1.54$~\AA~ neutrons is included in the same panel, which is largely consistent with the $\lambda = 2.41$~\AA~ data except for discrepancy at the lowest angles, which we attribute to increased background from the proximity to the direct beam at $\theta = 0$ in the $\lambda = 1.54$~\AA~ data.
The inset of Fig.~\ref{fig:hb2a}(b) shows NPD data under applied field where the short-ranged ice correlations partially give way to the magnetic Bragg peak at the  (2 0 0) position with $H = 0.2$~T, consistent with the partial recovery of remnant entropy over the same field range.

Most significant, however, are the solid curves in Fig.~\ref{fig:hb2a}(b)-(d), which represent the predicted scattering pattern associated with the spin configurations from the above MC consideration of heat capacity data.
Though the $T = 4$~K data is largely consistent with MC predictions, data at the lower two temperatures deviate significantly in the region $Q\approx0.55$~\AA$^{-1}$.
This is the (1 0 0) Bragg position, and can be interpreted as an excess of predicted $Q=X$ correlations driven by the sizable $J_{nnn}$ needed to reproduce the width of the heat capacity peak.
When using the same Hamiltonian to fit the NPD data instead (dashed curves), we get $J_{nnn} = 0$ which minimizes the $Q=X$ correlations. Not only are these fits poor, but as seen above in consideration of Fig.~\ref{fig:HC}, the inferred parameters lead to a systematic underestimation of heat capacity. We thus conclude that the next-nearest-neighbor DSI model is incapable of explaining the collective behavior of MgEr$_2$Se$_4$, leading to consideration of other effects.

\section{Discussion and conclusions}
In the context of the above data, it is important to consider the potential role of random fields due to local disorder, which have been suggested as a possible route to a QSL state in Pr pyrochlores\cite{martin2017, wen2017}.
We consider this explanation unlikely here, given the strong agreement between results on MgEr$_2$Se$_4$ and CdEr$_2$Se$_4$\cite{lago2010,gao2018}.
To address this possibility, however, we have measured magnetization and heat capacity of a second sample of MgEr$_2$Se$_4$, with a scattering pattern which implies significant structural disorder.
As shown in Appendix~\ref{appendix:disorder}, magnetization measurements on this sample were also well explained by our CEF calculations, implying local Ising physics.
However, heat capacity revealed a peak which is heavily skewed towards higher temperatures and, surprisingly, exhibits full remnant Pauling entropy.
Thus, it seems that disorder impedes, rather than encourages, the mechanisms leading to deviations from the DSI model.

We thus consider one last possibility: that significant quantum fluctuations are driven by transverse spin couplings.
On its face exotic, this option is in fact the least speculative, as quantum fluctuations have been predicted for Kramers doublets of the form we have observed\cite{onoda2011,huang2014,iwahara2015,rau2015,li2017}.
In contrast to the dominant dipolar character of moments in Dy$_2$Ti$_2$O$_7$ and Ho$_2$Ti$_2$O$_7$\cite{rau2015}, the current work and Ref.~\onlinecite{gao2018} show that moments in the Er-spinels contain a sizable ($\approx$1/3) leading order multipolar correction.
Such corrections should create transverse exchange couplings $J_\perp\propto \delta_i^2/\alpha^2$, where $\alpha$ and $\delta_i$ are the coefficients of the $\ket{15/2}$ and next leading order $\ket{J_z}$ term in the ground state wavefunction\cite{rau2015}.
Yb$_2$Ti$_2$O$_7$, known for complex quantum behavior from proximity to competing phases \cite{15_robert_Yb2Ti2O7,15_Jaubert_Yb2Ti2O7} including a quantum spin liquid state \cite{13_savary_QSL_Yb2Ti2O7_proximity}, has large transverse coupling experimentally determined as $J_\perp/J_z \approx 0.30$\cite{ross2011}.
This level of transverse coupling has a large effect on the material's properties\cite{changlani2017}.
Comparing to MgEr$_2$Se$_4$, our data implies a multipolar exchange that gives $J_\perp/J_z \approx 0.15$.
This is not a negligible effect, and should have immediately measurable consequences.
The anomalously fast monopole hopping rates recently reported for CdEr$_2$Se$_4$\cite{gao2018} may be one such example.
More direct confirmation may come from diffuse scattering measurements of single crystals, which would also facilitate tests of novel predictions for materials with DO doublets\cite{huang2014,li2017}.

Taken together, the collective data on MgEr$_2$Se$_4$ paint a picture of a spinel-based pyrochlore which provides an interesting counterpart to existing 227 oxides. The diffraction, heat capacity and inelastic neutron scattering results above leave very little doubt that this material contains the lattice, Ising anisotropy and ferromagnetic exchange necessary to drive spin ice behavior, and there is strong circumstantial evidence to infer significant quantum fluctuations. We further note that MgEr$_2$Se$_4$ is just one member of a series of magnesium rare earth selenides\cite{higo2017,reig2018}, some of which have an even larger capacity for quantum effects. These results, coupled with recent work on CdEr$_2$Se$_4$ and CdEr$_2$S$_4$\cite{gao2018}, may portend the opening up of a new class of magnetic spinel chalcogens, which can contribute meaningfully to the current research into pyrochlore materials.

\begin{acknowledgments}
The authors acknowledge useful discussions with P. Schiffer, G. Sala and G. Chen. This work was sponsored by the National Science Foundation, under grant number DMR-1455264-CAR. D.R. further acknowledges the partial support of by the U.S. D.O.E., Office of Science, Office of Workforce Development for Teachers and Scientists, Office of Science Graduate Student Research (SCGSR) program. The SCGSR program is administered by the Oak Ridge Institute for Science and Education for the DOE under contract number DE-AC05-06OR23100. Synthesis and thermodynamic measurements were carried out in the  Materials Research Laboratory Central Research Facilities, University of Illinois. Scattering measurements were conducted at the Center for Nanophase Materials Sciences, at the High Flux Isotope Reactor and at the Spallation Neutron Source, each DOE Office of Science User Facilities operated by the Oak Ridge National Laboratory. This work is part of the Blue Waters sustained petascale computing project, which is supported by the National Science Foundation (award numbers OCI-0725070 and ACI-1238993) and the State of Illinois.
\end{acknowledgments}



\appendix

\section{Monte Carlo simulation}
\label{appendix:MC}
We performed Monte Carlo simulations on a 2048-site cluster with periodic boundary conditions in all directions. Our Hamiltonian (Eq.~\ref{eq:hamiltonian}) includes nearest and next nearest neighbor Ising interactions and long range dipole-dipole interactions:

\begin{multline}
H = -3J_{nn}\sum_{\langle i,j \rangle}\mathbf{S}_i^{z_i}\cdot\mathbf{S}_j^{z_j}
-3J_{nnn}\sum_{\langle\langle i,j \rangle\rangle}\mathbf{S}_i^{z_i}\cdot\mathbf{S}_j^{z_j}\\
+ \frac{3D_{nn}}{5}r_{nn}^3\sum_{i<j}\left(\frac{\mathbf{S}_i^{z_i}\cdot\mathbf{S}_j^{z_j}}{|\mathbf{r}_{ij}|^3} - \frac{3(\mathbf{S}_i^{z_i}\cdot\mathbf{r}_{ij})(\mathbf{S}_j^{z_j}\cdot\mathbf{r}_{ij})}{|\mathbf{r}_{ij}|^5}\right),
\label{eq:hamiltonian}
\end{multline}

\begin{figure}[hbt]
 \centering
 \includegraphics[width=0.9\columnwidth]{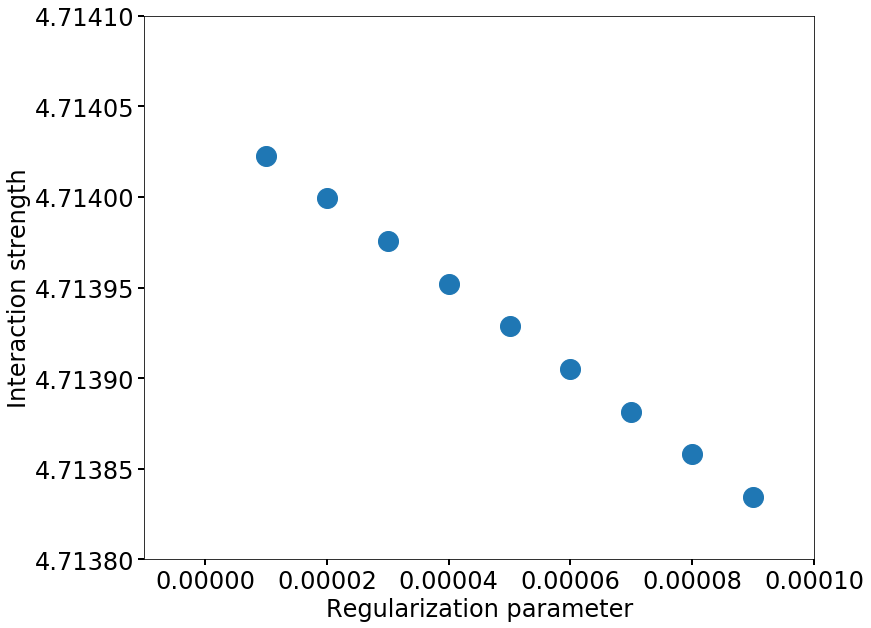}
 \caption{An example plot showing the numerical extrapolation of the interaction strength between an example pair of spins as the normalization parameter $s\rightarrow0$.}
 \label{appendix:MC_regulatization}
\end{figure}

To deal with the dipole term, one wants to sum over an infinite number of images.
Because the sum is conditionally convergent, the order in which this sum is taken affects the answer.
One approach to choosing the order of this sum is to use the Ewald technique.
An alternative approach (see section 4 of Ref.~\onlinecite{de1980simulation}), which we utilize, is to regularize the sum spherically by multiplying the contribution of the image which is in box $n=(n_x,n_y,n_z)$ by $\exp \left(-s\left|n\right|^2\right)$.
This regularization forces the sum to be absolutely convergent for all $s>0$.  We then numerically extrapolate to $s \rightarrow 0$ by evaluating the real space component from several values of finite $s$.
A sample extrapolation plot for a fixed $(i,j)$ is shown in Figure \ref{appendix:MC_regulatization}. Therefore both Ewald and this approach account for the long-range part of the interaction beyond simple truncation.
We benchmarked this method against the results obtained via Ewald summation \cite{denhertog2000} obtaining the same results. All data presented in the main text is obtained using the extrapolated parameters.

The specific heat,
\begin{equation}
C_{V}(T) = \frac{dE(T)}{dT}
\end{equation}
is computed by taking the derivative of this spline.
Spin configurations were fed into the program SPINVERT \cite{paddison2013} to obtain predicted powder-averaged diffuse neutron scattering patterns, $I(Q)$.

\begin{figure}[hbt]
 \centering
 \includegraphics[width=\columnwidth]{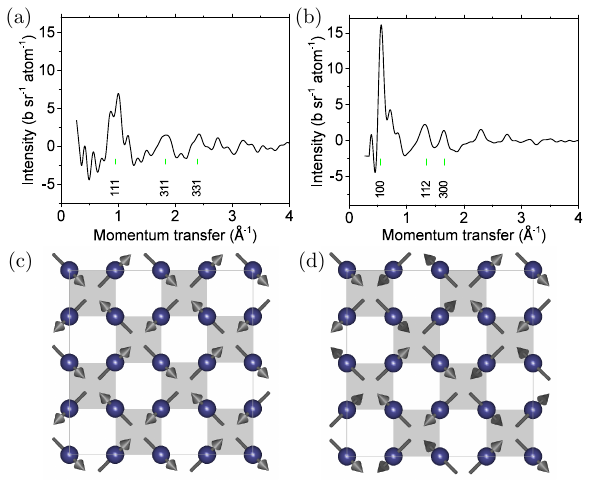}
 \caption{$I(Q)$ for simulated temperature subtracted NPD pattens for the case of ferromagnetic (a) and antiferromagnetic (b) $J_{nnn}$. Real space spin configurations for ferromagnetic (c) and antiferromagnetic (d) ordered phases as viewed along the cubic $[0 0 1]$ axis.}
 \label{fig:MC_order}
\end{figure}

For a given choice of $J_{nn}$, it was found that sufficiently large $|J_{nnn}|$ drove a transition into a long-range ordered state.
Figure~\ref{fig:MC_order} shows the neutron $I(Q)$ and real space pattern of spins for the case of both ferromagnetic and antiferromagnetic $J_{nnn}$.
Both ordered states preserve the two-in-two-out constraint of the spin-ice ground state.
The ferromagnetic state prefers a state which preserves the symmetries of the Fd$\bar{3}$m space-group, and thus demonstrated spin-spin correlations at locations consistent with a Q=0 state. Real space spin configurations indicate a similar state as preferred by Ho$_2$Ti$_2$O$_7$ and Dy$_2$Ti$_2$O$_7$ for applied fields $\mathbf{H} \parallel [0 0 1]$\cite{fennell2005}, or by ferrimagnetic spinels\cite{macdougall2012}.
The expected scattering pattern $I(Q)$ for this state is shown in Fig.~\ref{fig:MC_order}(a) and the corresponding magnetic structure is depicted in Fig.~\ref{fig:MC_order}(c).

The antiferromagnetic interaction case preferred a state which broke Fd$\bar{3}$m symmetry, and demonstrated distinct anti-correlations between chains of spins along the $[1 1 0]$ direction which are antiparallel to neighboring chains. Neutron intensity indicates a $Q=X$ phase, in that it shows correlations at the cubic $(1 0 0)$ and equivalent Bragg positions\cite{harris1997}. Although similar to the $Q=X$ phase favored by $\mathbf{H} \parallel [1 1 0]$ fields in that both have antiparallel chains of spins along the $[1 1 0]$ direction\cite{fennell2005}, the current phase is actually distinct, in that there is no net moment. These plots are shown in Fig.~\ref{fig:MC_order}(b) and (d) for $I(Q)$ and real space respectively.

\section{Magnetic Properties of the Impurity Phase}\label{appendix:mag_impurity}

\begin{figure}[hbt]
 \centering
 \includegraphics[width=\columnwidth]{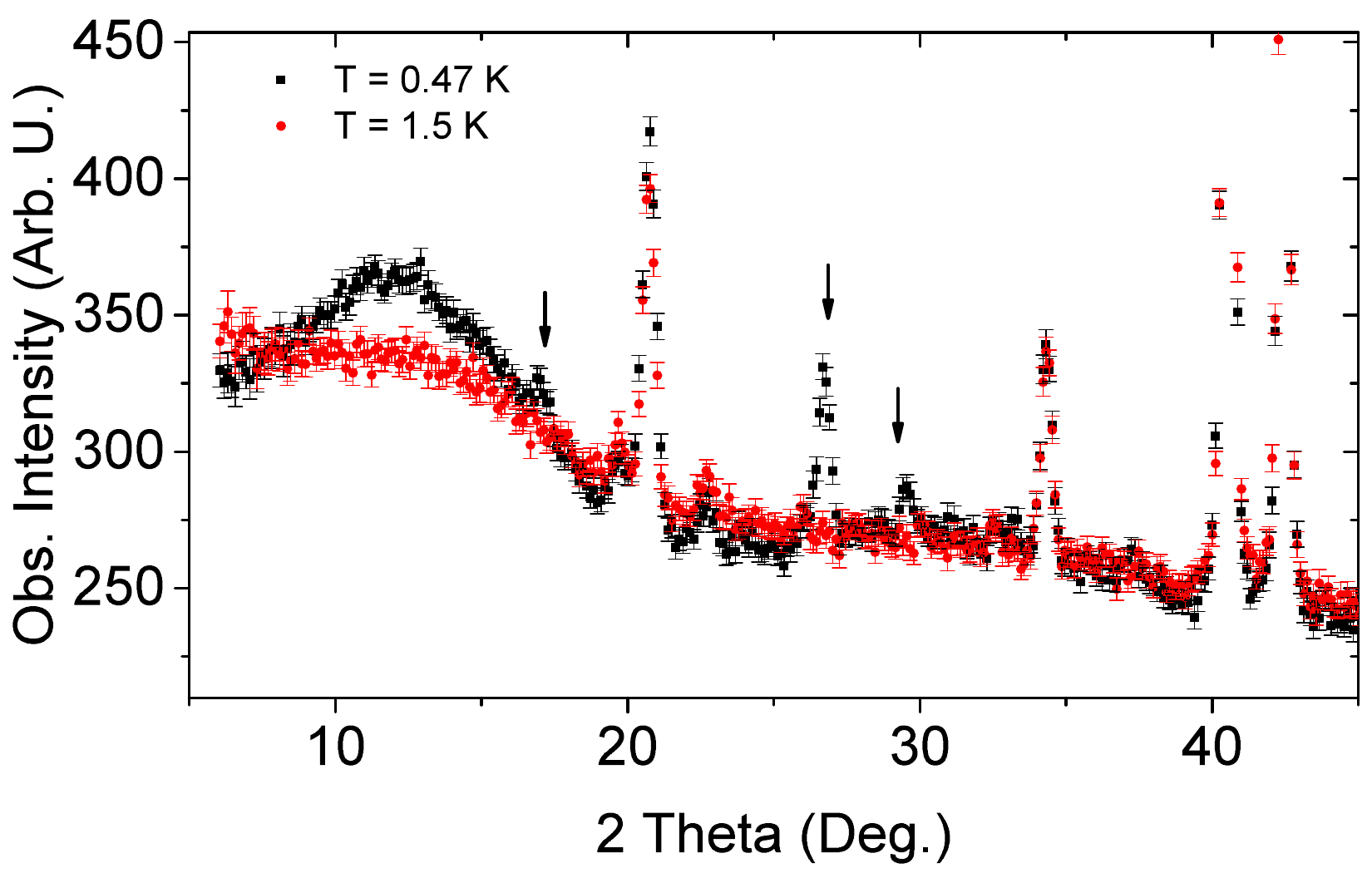}
 \caption{NPD data at 1.5 K and 470 mK, with magnetic impurity peaks marked by arrows.}
 \label{fig:NPD_impurity}
\end{figure}

In Fig.~\ref{fig:NPD_impurity} we show the low temperature NPD patterns at $T = 1.5$~K and 0.47~K, without the $T = 38 $~K pattern subtracted.
In addition to broad features associated with spin-ice correlations and Bragg peaks associated with the lattice, the data shows a series of weak Bragg magnetic peaks which appear only at the lowest temperature.
These peaks were not indexable in the Fd$\bar{3}$m space group of the spinel structure and had total weight of $0.55(5)~\%$ of the diffuse correlations.
We thus associate them with the same impurity phase discussed in the main text and we didn't include data points at those peaks in Fig.~\ref{fig:hb2a}(b), for cosmetic purposes only.
Neither the position nor the weight of the magnetic impurity peaks are capable of accounting for the large peak predicted by MC simulations using best fit parameters for heat capacity data, and the presence or absence of these peaks in the NPD data do not change the conclusion of this work in any way. We estimated the potential contribution to the heat capacity due to the onset of spin order in a $0.55(5)~\%$ impurity phase, and incorporated this value into the error bars when determining magnetic heat capacity in the main text.


\section{INS fitting method and results}\label{appendix:INS_fit}

The resulting wavefunctions for all CEF levels are shown in Table~\ref{tab:CEF}. The code for fitting the INS data to the CEF model was written in MATLAB.
The assumptions for the refinement are as follows: the CEF levels are thermally populated according to the partition function $Z = \sum_n \text{exp}\left( -\beta E_n \right) $, where $E_n$ is the energy of the $n^{\text{th}}$ CEF level.
Excitation energies are determined by diagonalizing the Hamiltonian
\begin{equation}
  H = \sum_{nm} B_n^m O_n^m,
  \label{eq_ham2}
\end{equation}
where $B_n^m$ are the crystal field parameters and $O_n^m$ are Steven's operators.
In order to isolate the unique solution where the wavefunctions are maximally parallel/antiparallel to the $\langle 1 1 1 \rangle$ directions, a small field of $10^{-7}$ T along the $\langle 111 \rangle$ is added to the potential; we note that this field is too small to change the energy of the calculated CEF levels.
Peak intensities are given by $p_n \bra{\psi_n} J_{\alpha} \ket{\psi_m}^2 $ where $p_n$ is the probability of an Er$^{3+}$ ion being in the $\psi_n$ state, and $J_\alpha = J_- + J_+ + J_z$.
The fitting was done using a random grid search method that explores the six dimensional phase space of the six non-zero $B_n^m$ coefficients.
The variation within this phase space was done by projecting random vectors in the phase space and then finding the least squares minima of the simulated pattern along those vectors.
The overall minimum is then taken as the next starting point, and the process is repeated until convergence.
The program was initially run with the lowest-lying CEF excitation fixed to $E = 4.1$~meV, thereby limiting the search to vectors in the five dimensional manifold that satisfied this condition.
After initial convergence, this condition was relaxed.
To check against false minima, the program was run eight separate times, and it was confirmed to converge to the same values.
\par
In order to find errors of the $B_n^m$ CEF parameters we use a gradient method which is considerably faster than the grid search method, although it is less robust against false minima.
For each $B_n^m$ we fix the parameter's value and minimize $\chi^2$ by moving along the gradient in the remaining five non-fixed CEF parameters.
This is repeated for $B_n^m$ fixed at a value progressively farther from the minima until $\chi^2$ has increased by one, which we define as the upper and lower bounds of the error.
By doing this for all of the parameters we get the error bars for each CEF parameter.

\begin{table*}[bht] 

\begin{ruledtabular}
   \begin{tabular}{ l l l p{0.75\linewidth} }
    \multicolumn{4}{c}{MgEr$_2$Se$_4$ crystal field levels} \tabularnewline
    \hline
    $n$ & $\delta E$ & $I_{relative}$ &  $\Psi_{n}$  \tabularnewline
    \hline
    0 & 0 & - & $ \pm0.9165(7)\ket{\pm15/2} + 0.360(1)\ket{\pm9/2}
 \pm 0.158(2)\ket{\pm3/2} -  0.073(2)\ket{\mp3/2}
 \pm 0.0036(7)\ket{\mp9/2} + 0.0035(14)\ket{\mp15/2}$  \tabularnewline
    1 & 4.155 & 1 & $ 0.734(5) \ket{\pm13/2} \mp 0.488(1) \ket{\pm7/2} +
    0.435(10) \ket{\pm1/2} \pm 0.177(2) \ket{\mp5/2} + 0.0504(6) \ket{\mp11/2}$ \tabularnewline
    2 & 6.279 & 0.0517 & $ \pm 0.480(9) \ket{\pm13/2} - 0.071(5) \ket{\pm7/2} \mp 0.868(5) \ket{\pm1/2}
    - 0.072(3)  \ket{\mp5/2}  \pm 0.066(4) \ket{\mp11/2}$ \tabularnewline
    3 & 9.193 & 0.0862 & $\pm 0.267(8) \ket{\pm15/2} + 0.252(5) \ket{\pm9/2} \mp 0.918(16) \ket{\pm3/2} - 0.13(11) \ket{\mp3/2} \pm 0.040(27) \ket{\mp9/2}  - 0.068(30) \ket{\mp15/2} $ \tabularnewline
    4 & 10.133 & 0.2846 & $\pm 0.6651(9) \ket{\pm11/2} - 0.7238(8) \ket{\pm5/2} \mp 0.097(4) \ket{\mp1/2} - 0.0107(7) \ket{\mp7/2} \mp 0.156(4) \ket{\mp13/2} $ \tabularnewline
    5 & 27.273 & 0.0329 & $\mp0.692(1) \ket{\pm11/2} - 0.571(2) \ket{\pm5/2} \mp 0.187(2) \ket{\mp1/2} + 0.342(3) \ket{\mp7/2} \mp 0.207(2) \ket{\mp13/2}$ \tabularnewline
    6 & 29.91 & 0.0188 & $ + 0.290(1) \ket{\pm15/2} \pm 0.8977(4) \ket{\pm9/2} + 0.3155(15) \ket{\pm3/2} \pm 0.102(2) \ket{\mp3/2} + 0.011(3) \ket{\mp9/2} \mp 0.0014(9) \ket{\mp15/2} $ \tabularnewline
    7 & 29.945 & 0.0055 & $\pm0.4035(14) \ket{\pm13/2} + 0.7995(14) \ket{\pm7/2} \pm 0.1098(10) \ket{\pm1/2} + 0.3437(3) \ket{\mp5/2} \mp 0.269(2) \ket{\mp11/2}$ \tabularnewline
\end{tabular}
\end{ruledtabular}
 \caption{The full CEF scheme of MgEr$_2$Se$_4$ as calculated from the best fit to the data. The energy levels, relative neutron scattering intensity at 0 K, and wavefunctions are presented for the 8 CEF doublets associated with the ground state manifold.}
 \label{tab:CEF}
\end{table*}

\section{Magnetization calculation}\label{appendix:mag_calc}
Magnetization curves in the main text were obtained through calculations which took into account the full CEF Hamiltonian for the $J = \frac{15}{2}$ multiplet of Er$^{+3}$.
This method allowed us to describe the moment of the material at both higher temperatures, where multiple CEF levels are occupied, and at higher fields, where mixing of the states leads to an increased moment.

In order to calculate the moment, we ignored interactions between moments and treated the problem in the single ion picture.
The total Hamiltonian is thus the CEF plus the Zeeman energies
\begin{equation}
H = H_\text{CEF} + H_Z.
\label{eq:in_field_potential_appendix}
\end{equation}
The CEF Hamiltonian is defined as before in Eq.~\ref{eq:CEF}
and the Zeeman term
$H_Z = - g_J \mathbf{H} \cdot \hat J $
where $g_J$ is the Land\'e g factor for Er$^{3+}$.
The combined Hamiltonian was diagonalized to give the energies $E_n$ and wavefunctions $\psi_n$ which were used to find the partition function $Z = \sum_n \exp(- \frac{E_n}{k_b T})$ and the contribution to the moment
\begin{equation}
M_n = \frac{\bra{\psi_n}\mathbf{H} \cdot \hat{J} \ket{ \psi_n }}{\left|\mathbf{H}\right|}.
\end{equation}
We then powder averaged by integrating over the polar angle $\theta$
\begin{equation}
\int_0^\frac{\pi}{2} g_l \sum_n p_n(\mathbf{H},\theta) M_n(\mathbf{H},\theta) \sin(\theta) d\theta,
\end{equation}
where $p_n$ is simply $\frac{\exp(- \frac{E_n}{k_b T}) }{Z}$.
Due to symmetry we did not need to average over the azimuthal angle, and only needed to integrate to $\frac{\pi}{2}$.
This gave the full powder averaged magnetization per Er atom including the effects of the mixing of higher energy CEF doublets.
\par
\begin{figure}[hbt] 
 \centering
\includegraphics[width=\columnwidth]{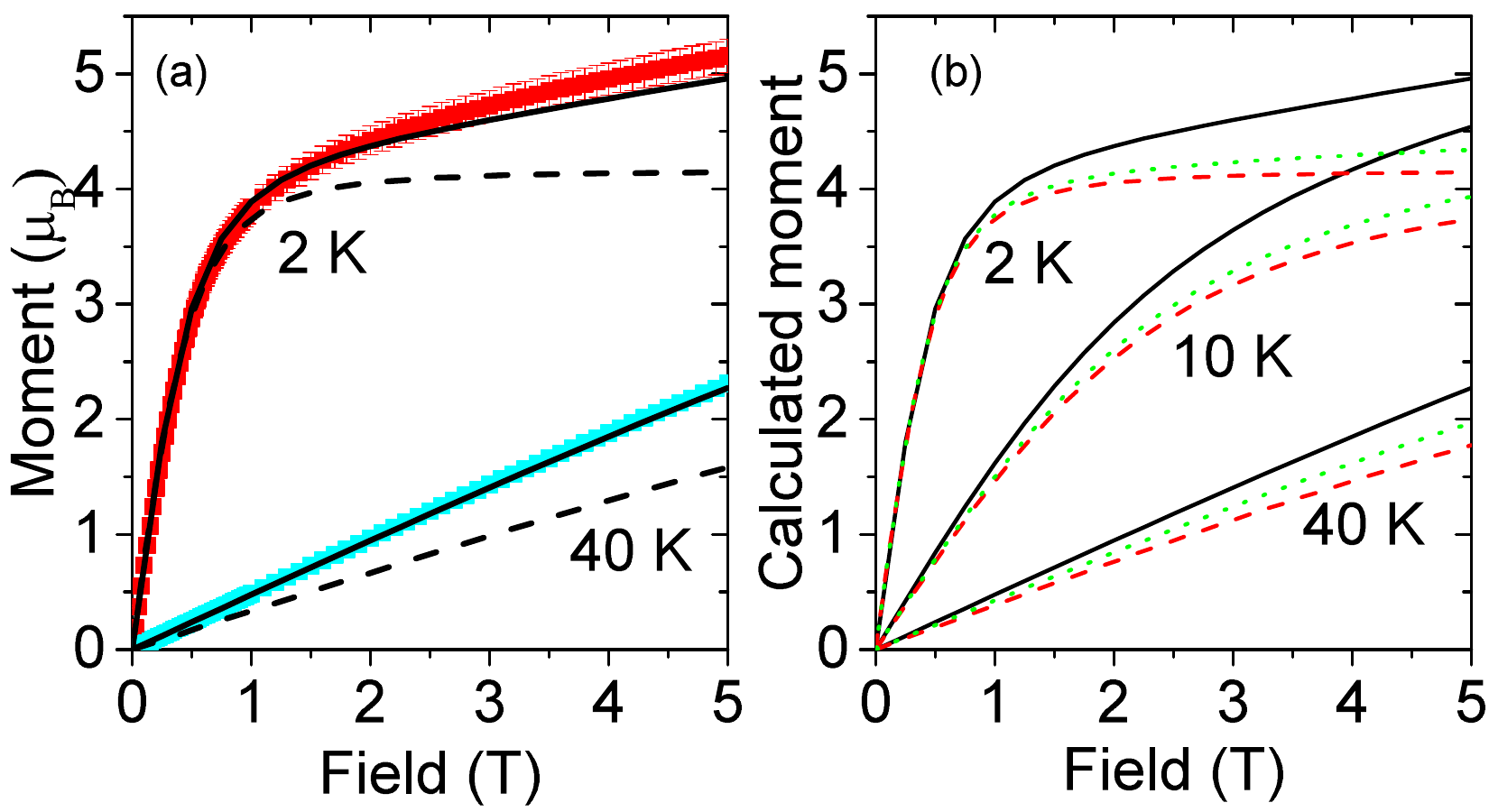}
\caption{Plots showing the importance of including the contributions from excited CEF levels in the calculation of the magnetization for MgEr$_2$Se$_4$. The left panel presents the measured data at 2 and 40~K as red and blue marks respectively. The calculated magnetization is superimposed on the data for the cases where only the ground state wavefunctions are considered (dashed curves) and when contributions from higher-energy wavefunctions are included (solid curves). The right panel compares our exact moment calculation (black) to predictions of the effective spin-1/2 model (green), which neglects perturbation effects and effects of higher energy CEF levels. The red curve shows the exact calculation again, but where excited CEF levels 15.845~meV artificially added to their energy, showing the presence of these modes to be the dominant effect.}
 \label{fig:mag_comparison}
\end{figure}

The effects of higher energy wavefunctions were found to be particularly important to the calculation of magnetization. To highlight this fact, Fig. \ref{fig:mag_comparison} shows a comparison of the data for this material calculated with and without these effects considered.

At any fields higher than 1 T, the difference between the calculated magnetization and the magnetization from a simple two level model is considerable.
In Fig.~\ref{fig:mag_comparison}(b) the same calculation is repeated but with $+15.845$~meV artificially added to all of the excited energy levels, in order to bring the overall first excited energy to 20 meV. This value is much closer to the energy of the 227 rare earth pyrochlores studied, and we can see that the effect now becomes far less important; the resulting curves are quite similar to the moment calculated considering only the ground state doublet.

\section{The Effects of Disorder}\label{appendix:disorder}

\begin{figure}[!hbt]
 \centering
 \includegraphics[width=\columnwidth]{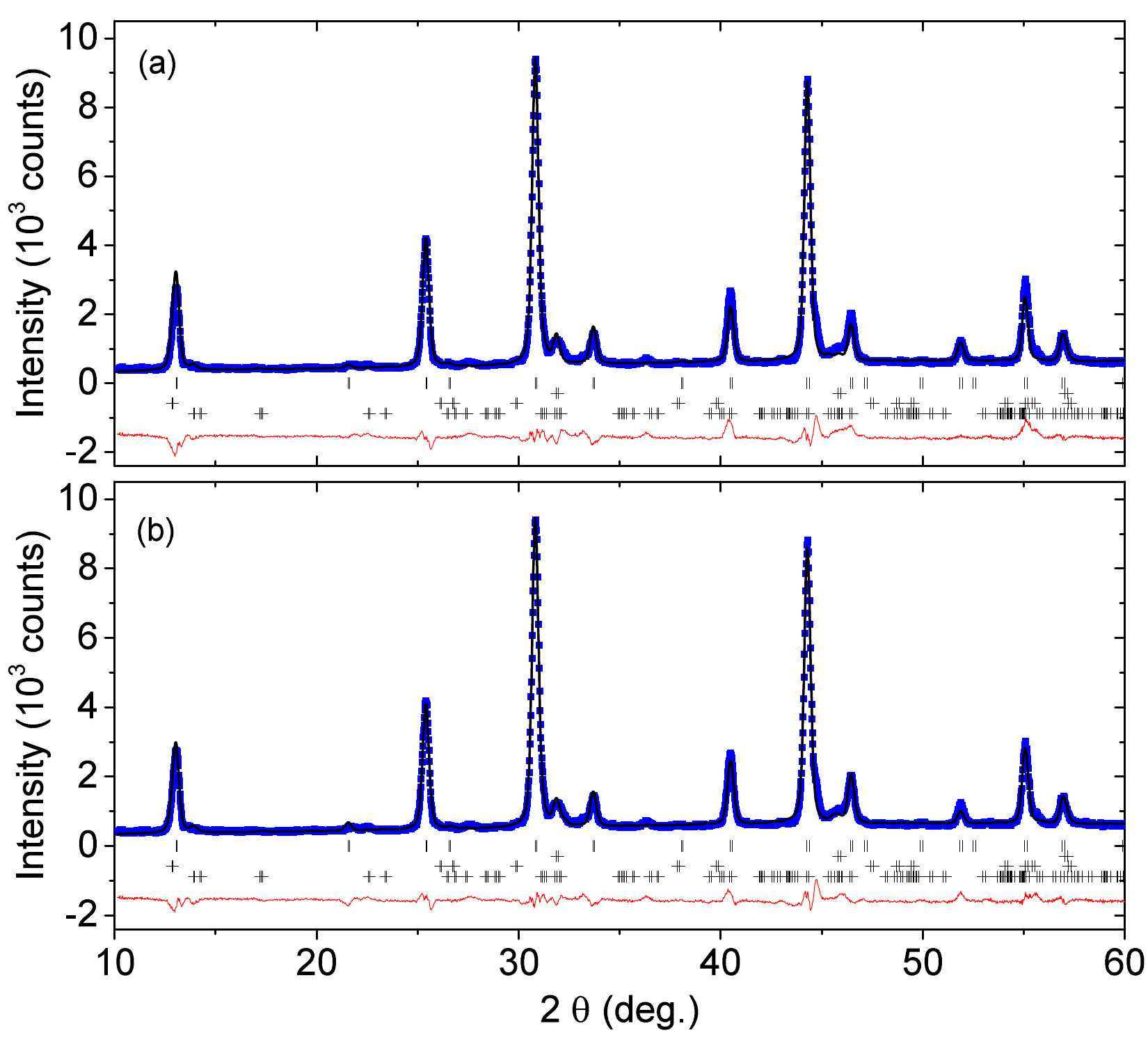}
 \caption{X-ray diffraction pattern from the disordered sample-B MgEr$_2$Se$_4$, with fits performed using a best-fit refinement to the spinel structure (a) and using a model-independent Le Bail analysis (b).}
 \label{fig:sampleB_XRD}
\end{figure}

In order to explore the effects of disorder on the material properties of MgEr$_2$Se$_4$, we performed a series of thermodynamic measurements on a second, less pure powder sample with a demonstrably higher level of local disorder. In Fig.~\ref{fig:sampleB_XRD}(a) we show the XRD pattern from this sample (which we call ``sample-B''),  fit using a standard Rietveld refinement ($\chi^2 = 5.7$), while Fig.~\ref{fig:sampleB_XRD}(b) presents the results of Le Bail analysis on the same dataset ($\chi^2 = 10.6$).
As compared to our primary sample (henceforth referred to as ``sample-A''), the XRD data on sample-B revealed a marginally higher fraction of impurities ($\approx 10 \%$ total), and a Le Bail fit which improved $\chi^2$ considerably over refinement values.
Though the peak positions are consistent with a cubic Fd$\bar{3}$m space group, the inability of the standard refinement to describe peak heights within the spinel model, even allowing for variations in site occupancy, reveals the presence of a significant level of structural disorder.

\begin{figure}[!hbt]
 \centering
 \includegraphics[width=\columnwidth]{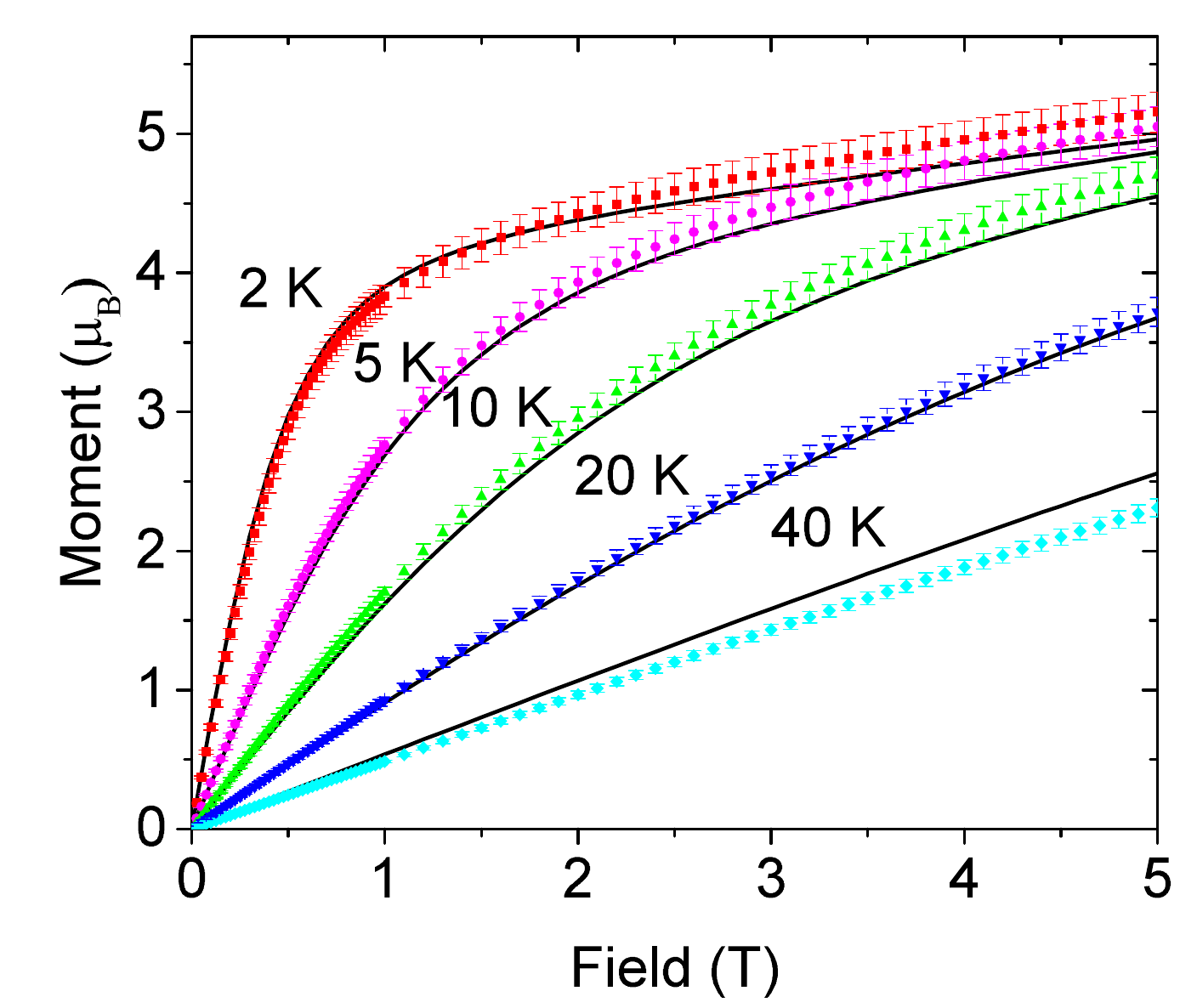}
 \caption{Magnetization of sample-B versus field at several temperatures, compared to the predictions based on the CEF level scheme determined from INS data for sample-A. The strong agreement here confirms that spins in MgEr$_2$Se$_4$ remain strongly Ising-like, even in the presence of significant disorder.}
 \label{fig:sampleB_magnetization}
\end{figure}

Figure~\ref{fig:sampleB_magnetization} shows the magnetization of sample-B over a range of fields and temperatures, which are interesting to compare to measurements on sample-A presented in the main text.
Although INS measurements were not performed on sample-B, we assume a similar CEF to calculate expected magnetization curves shown as solid lines in Fig.~\ref{fig:sampleB_magnetization}.
Despite not having a separate INS study of the CEF levels of sample-B we find that this agreement in magnetization data shows that the CEF is not significantly modified by disorder, as would be expected for a Kramers ion.

\begin{figure}[!hbt]
 \centering
 \includegraphics[width=\columnwidth]{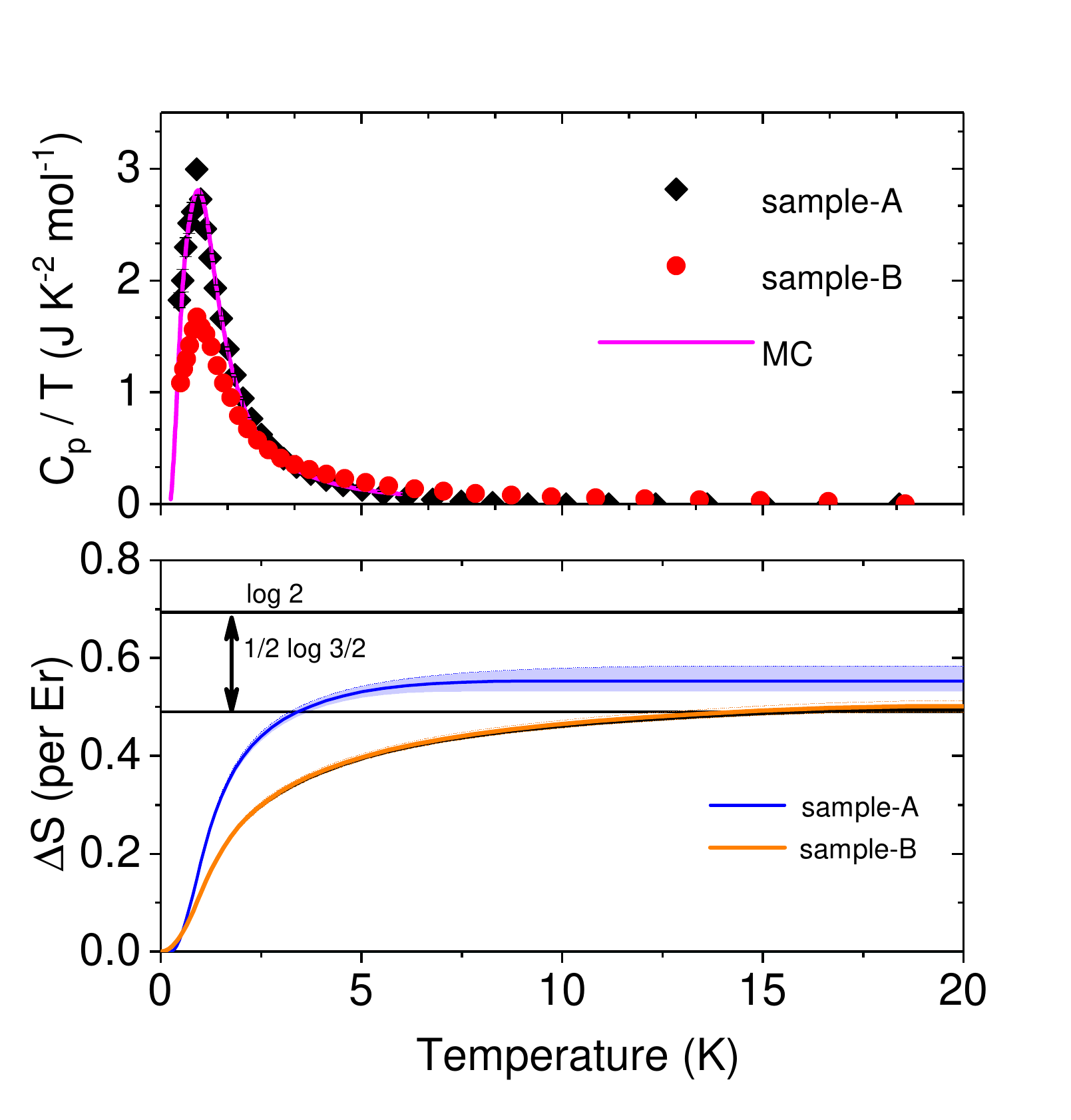}
 \caption{A comparison of the magnetic heat capacity (top) and associated entropy (bottom) of high quality and disordered powder samples of MgEr$_2$Se$_4$.}
 \label{fig:sampleB_HC}
\end{figure}

In contrast, heat capacity is modified significantly by disorder effects, as revealed by Fig.~\ref{fig:sampleB_HC}. In Fig.~\ref{fig:sampleB_HC}(a), we show a comparison of the heat capacity of the two samples, again with the best fit MC curve for sample-A. The peak in the heat capacity for sample-B is reduced by almost a factor of two, and we have confirmed that there is no spin configuration in the next-nearest-neighbor dipole-ice model capable of reproducing this behavior. Further inspection reveals that the heat capacity of sample-B is not uniformly reduced, but rather that the peak is skewed to higher temperature. This leads to a long high-temperature tail wherein the curve for sample-B lies above sample-A. Integrating the area under these curves leads to the associated entropy curves in Fig.~\ref{fig:sampleB_HC}(b). Quite surprisingly, we find that sample-B recovers full Pauling residual entropy as $T\rightarrow0$~K, in direct contrast to the conclusions of the main text for sample-A. This suggests that disorder acts to hinder the mechanisms which are leading to the reduction of residual entropy in relatively pure samples of MgEr$_2$Se$_4$.

\pagebreak
\pagebreak
\par


%

\end{document}